\documentstyle[prl,aps,epsf]{revtex}

\begin{document}
\draft
\widetext
\twocolumn[{
\title{Optical excitations in a non-ideal Bose gas}
\author{M.~\"O.~Oktel, L.~S.~Levitov}
\today
\address{
12-112, Massachusetts Institute of Technology, Cambridge MA 02139
}
\maketitle

\mediumtext
\begin{abstract}
  Optical excitations in a Bose gas are demonstrated to be very
sensitive to many-body effects. At low temperature the momentum
relaxation is provided by momentum exchange collisions, rather
than by elastic collisions. A collective excitation mode forms,
which in a Boltzmann gas is manifest in a collision shift and
dramatic narrowing of spectral lines.
  In the BEC state, each spectral line splits into two
components. The doubling of the optical excitations results from
the physics analogous to that of the second sound. We
present a theory of the line doubling, and calculate the
oscillator strengths and linewidth.
   \end{abstract}
\pacs{PACS:                   }

}]
\narrowtext

Recent experiments on the Bose-Einstein condensation in
magnetically trapped gases\cite{BEC1,BEC2,BEC3,BEC_H} stimulated
studies of many-body phenomena in such systems. Collective
excitations of several kinds\cite{HydroResponse}
have been investigated, including
density waves\cite{BECsound} and the second sound\cite{2-sound},
excited by shaking the trapped BEC samples, or by an abrupt
change of the interaction between
particles\cite{BECinteraction}. All these responses are of
hydrodynamical character, with the
excitation frequencies being not much greater than the trap
frequency, and the wavelengths of order of the sample dimension.

Optical excitations are also
sensitive to the nature of
quasiparticles and interactions between them. However,
quasiparticles in a weakly non-ideal Bose gas, according to
Bogoliubov theory, differ from free particles
only at very low temperatures\cite{light_scattering}:
 $\sim T_{\rm B}=(4\pi\hbar^2/m)aN\ll T_{\rm BEC}$, where $N$
is the gas density, and $a$ is the {\it s}--wave scattering
amplitude. In this letter we describe the many--body phenomena
resulting from a change of the particle internal state due to
optical transition. We shall see that these many--body effects can be
significant even at temperatures as high as $T\sim T_{\rm
BEC}$, such that the difference of the quasiparticles and
particles plays no role.

The reason for the optical excitations being special is that the
excited particle spends a long time in the state which is a
coherent superposition of the ground state and the excited
state. This time, which is of order of the inverse spectral line
width, can be very long, so that the excited particle, while
being in the superposition state, interacts with many other
particles. The collisions of particles in different but {\it
non-orthogonal} internal states have special properties, first
discussed in the context of the problem of spin excitations in
gases\cite{Bashkin,Laloe,Verhaar} (see also the work on
collision frequency shifts in cesium fountains\cite{cesium}).

Let us consider a gas of identical atoms, all in the ground
internal state $|1\rangle$. Suppose that one of the atoms is transferred
to a superposition
state $|1'\rangle$, having finite overlap with the ground
state:
$|1'\rangle=\alpha |2\rangle +\beta |1\rangle$.
Generally, there are three different kinds of
collisions between this atom and other atoms:
(i) ordinary elastic
collisions, occurring with the frequency $\tau^{-1}_{\rm
el}=v_Ta^2_{12}N$, where $v_T=(2T/m)^{1/2}$ is thermal velocity,
and $a_{12}$ is the scattering length;
(ii) forward collisions, in which the atoms interact without
changing direction and velocity of motion; (iii) momentum
exchange collisions, in which interacting atoms switch their
trajectories and velocities. The forward and momentum exchange
scattering rates are both given by $\tau^{-1}_{\rm exch}=4\pi\hbar a_{12}N/m$,
{\it i.e.}
are temperature independent. Thus, at temperatures
$T\le\hbar^2/ma^2_{12}$ the collisions (ii) and (iii) dominate.

In this regime, named {\it quantum gas} by
Bashkin\cite{Bashkin}, collective effects appear due to
collisions of type (iii). Note that forward collisions (ii)
affect the non-interacting particle picture only to the extent
that they correct external potential seen by a particle by an
amount proportional to the density of other particles. The role
of the collisions (iii) is quite different. They provide
mechanism of momentum transfer between different particles with
the rate $\tau^{-1}_{\rm exch}$ much higher than
that of energy relaxation. Thus, at times $\le\tau_{\rm el}$
the particles can exchange momenta many times without changing
the overall energy distribution. As a result, all excitations in
the system with wavelength greater than $v_T\tau_{\rm exch}$,
the path between momentum exchange collisions, become collective
modes.

The  collective  phenomena in optical excitations arising due to
momentum exchange collisions differ from those in the spin--wave
dynamics\cite{Bashkin,Laloe,Ruckenstein}.  The  reason  is  that   the
interactions between atoms in different orbital states depend on
the states, while for the atoms in different spin states all the
interactions are the same.

\noindent{\it Problem:} We consider optical excitations in
a Bose gas of uniform density $N$. Each atom of the gas can be in
the ground state or in the excited state
(denoted  by  $1$  and  $2$,  respectively). The interaction
constants  $\lambda_{ij}$, $i,j=1,2$ are related by the
factor $4\pi\hbar^2/m$ with the $s-$wave
scattering lengths $a_{ij}$ ($a_{11}=a$).

The temperature range of
interest, $T\le \hbar^2/ma^2$, includes temperatures both above
and below $T_{\rm BEC}$. For the collective effects to be
pronounced, the excitation wavevector $k$ should be
$\le(v_T\tau_{\rm exch})^{-1}$, which means that
$c|k|\ll\omega_0$, the excitation frequency. Thus, optimal
situation is that of a two-photon excitation with the photons'
momenta equal and opposite.

For example, this is realized in
the recent study of the BEC in spin-polarized hydrogen by using
the $1S-2S$ Doppler free transition\cite{1S2S}.
In this experiment $k\simeq
2\pi/w$, where $w$ is the width of the laser beam used to excite
the $1S-2S$ transition.

Below we study the excitation above and below $T_{\rm BEC}$.
We calculate the frequency shift and the  linewidth of
the excitation. In the analysis we ignore spin degrees
of freedom of the gas, as well as the finite size of the gas
sample  confined  in  the  trap.  Thus, our results apply to the
excitations probed near the central region of  a  shallow  trap,
such that the density is nearly uniform.
Also, our treatment is restricted
to not very low temperatures, $T>\hbar/\tau_{\rm exch}\sim T_{\rm B}$,
where one can ignore the difference between free particles and
Bogoliubov quasiparticles.

\begin{figure}
\epsfxsize=75mm
\epsfbox{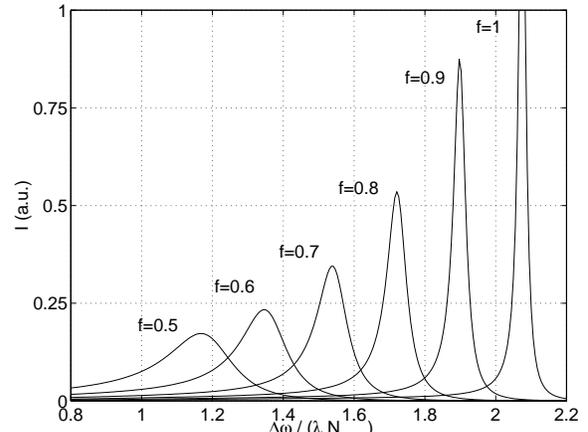}
\caption[FIG1.]{
\label{fig1}
Absorption spectrum (\ref{S(w)}),(\ref{P0})
above BEC at fixed temperature
and varying density: $N=fN_{\rm BEC}$, $0.5\le f<1$.
The frequency shift $ \Delta \omega $ is defined relative to the
that of a free atom at rest:
$\Delta\omega = \omega - \omega_0 - k^2/2m$. Spectral power ${\cal I}$
is normalized by particle density $N$. The excitation wavevector $k$ is
$0.5$ in the units of $\lambda_{12} N_{\rm BEC}/v_T$. The interaction constant
$\lambda_{11}$ is chosen to be $0$. Note that the peak position
follows the relation (\ref{dispersion}).
  }
\end{figure}

\noindent{\it Summary of results:} We show that the frequency of the excitation
is shifted from that
of a single atom by an amount {\it different} from the
mean density shift. The reason for that is that the momentum
exchange process couples different states degenerate in
energy, and therefore, one has to apply degenerate perturbation
theory to the many-body states to determine the correct frequency
shift.  For the Boltzmann
gas,  $T>T_{\rm  BEC}$, by   using
the   many-body   version   of  the  degenerate
perturbation theory (the   random   phase
approximation), we find the excitation
frequency:
  \begin{equation}\label{dispersion}
\omega_{1\to 2}(k)=\omega_0+2(\lambda_{12}-\lambda_{11})N
+\frac{v_T^2 k^2}{3\lambda_{12}N}\ ,
  \end{equation}
where the transferred momentum is assumed to be small in the
sense discussed above: $|k|<\lambda_{12}N/v_T$. The resonance
frequency at $k=0$ varies linearly with density (see
Fig.\ref{fig1}). Such a behavior is observed in the
experiment\cite{1S2S}, and the magnitude of the shift is found
to be in agreement with (\ref{dispersion}).

The linewidth of the excitation (\ref{dispersion}) due to momentum
exchange collisions is found to be
  \begin{equation}\label{linewidth}
\gamma_{\rm exch}
\simeq kv_T\ e^{-(\lambda_{12}N/kv_T)^2}\ .
  \end{equation}
(Elastic collisions contribute $\gamma_{\rm
el}\simeq\tau^{-1}_{\rm el}\ll\gamma_{\rm exch}$.) The abrupt
narrowing (\ref{linewidth}) of the spectral line at high densities
   \begin{equation}\label{DickeCriterion}
N>kv_T/\lambda_{12}
   \end{equation}
is similar to the motional (Dicke) narrowing effect (see
Fig.\ref{fig1}) This is because each momentum exchange collision
fully randomizes velocities, and so in the range
(\ref{DickeCriterion}) the collisions path, $v_T/\lambda_{12}N$,
becomes smaller than the wavelength $2\pi/k$. Another way to put
it is to draw an analogy with M\"ossbauer effect, i.e., to say
that in the regime (\ref{DickeCriterion}) the recoil momentum is
transferred collectively to all of the
atoms interacting with the excited atom while it is in phase
with the excitation field.

Below $T_{\rm BEC}$, the main change in the spectrum
is that the spectral line
splits into two components, both shifted from the single particle
frequency
$\omega_0$. At $k=0$ the frequencies are given by
   \begin{eqnarray}\label{2frequencies}
& & \omega_{1,2}=\omega_0+(\lambda_{12}-2\lambda_{11})N+X_{1,2}\ , \cr
& & (X_{1,2}-\lambda_{11}N_c)(X_{1,2}-\lambda_{12}N_T) =\lambda_{12}^2N_cN_T\ ,
   \end{eqnarray}
where $N_c$ is the condensate density, and $N_T=N-N_c$ is the thermal density.
The origin of the two frequencies (\ref{2frequencies})
is that in the presence of the condensate the excited atom can be in two distinct states,
``correlated'' and ``uncorrelated''. In the correlated state the kinetic
energy of excited atom is
$\sim T$, and so it participates in momentum exchange collisions which shift
$\omega$ as discussed above. In the uncorrelated state, the velocity of the
excited atom is zero, it does not take part in momentum exchange
processes, and thus does not acquire the additional energy shift.
Besides that, there are momentum exchange collisions which
transfer the atom from the correlated to the uncorrelated state and backwards
(see discussion below).
The difference of the two frequencies (\ref{2frequencies})
gives the frequency of coherent oscillations between the two states.

\begin{figure}
\epsfxsize=75mm
\epsfbox{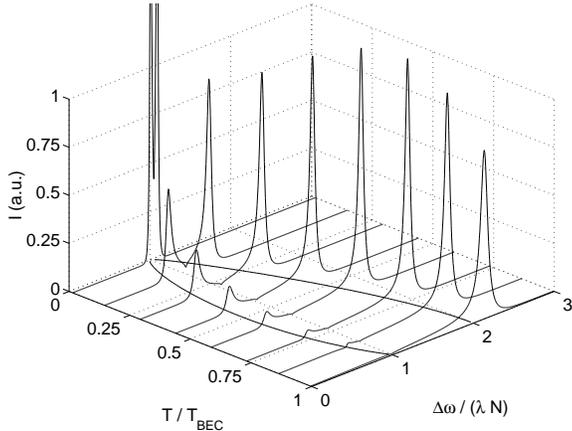}
\caption[FIG2.]{
\label{fig2}
Absorption spectrum (\ref{S(w)}),(\ref{P1})
in the BEC regime, shown for temperature
varying between $0$ and $T_{\rm BEC}$; density $N$ fixed. The
excitation wavevector $k$ is $2/3$ in the units of $\lambda_{12}N/v_T$.
Lines in the base plane indicate the peak positions (\ref{2frequencies})
for $k=0$. Note
narrowing of the spectral line with decreasing $T$, and strengthening of
the condensate peak due to increasing condensate fraction. (The
frequency shift $\Delta\omega$ is defined
in Fig.\protect\ref{fig1}; $\lambda_{11}=0$.)
  }
\end{figure}

\begin{figure}
\epsfxsize=75mm
\epsfbox{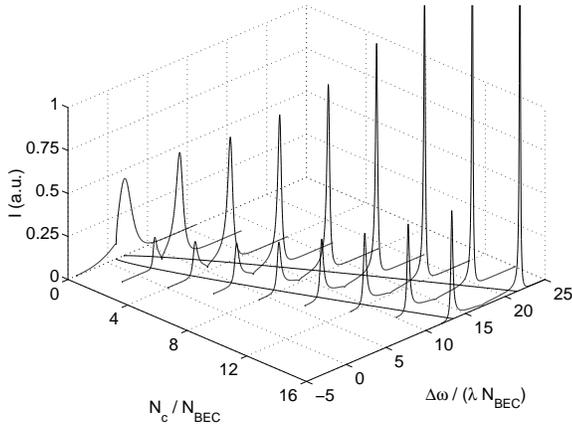}
\caption[FIG3.]{
\label{fig3}
Absorption spectrum (\ref{S(w)}),(\ref{P1}) in the BEC regime at fixed
temperature and density varying from $N_{\rm BEC}$ and up. Excitation wavevector $k=2.5$ in the units
of $\lambda_{12} N_{\rm BEC}/v_T$. Increasing condensate density leads to narrowing of
the peaks and to strengthening of the condensate
peak, as in Fig.\protect\ref{fig1}. ($\Delta\omega$
and ${\cal I}$ are defined in Fig.\ref{fig1}; $\lambda_{11}=0$.)
  }
\end{figure}

The behavior of the dispersion and the linewidth due to finite $k$
as a function of temperature and density is qualitatively similar to that
above $T_{\rm BEC}$ (see Figs.\ref{fig2},\ref{fig3}).

\noindent{\it Calculation}
The system is described by the
Hamiltonian ${\cal H}={\cal H}_1+{\cal H}_2+{\cal H}_{ext}$:
   \begin{eqnarray}
{\cal H}_1 &=& \sum\limits_p\
\frac{p^2}{2m}a^+_pa_p +
\sum\limits_{p_1+p_3=p_2+p_4}
\frac{1}{2}\lambda_{11}
a^+_{p_1} a_{p_2} a^+_{p_3} a_{p_4} \ \cr
  {\cal H}_2 &=& \sum\limits_p\
\left(\frac{p^2}{2m}+\omega_0\right)b^+_pb_p +
\sum\limits_{p_1+p_3=p_2+p_4}
\lambda_{12}
a^+_{p_1} a_{p_2} b^+_{p_3} b_{p_4} \ \cr
{\cal H}_{ext} &=& \sum\limits_{p,k} A_k e^{-i\omega t} b^+_{p+k}a_p \ +\ {\rm h.c.}
   \end{eqnarray}
where $a_p$ and $b_p$ are canonical Bose operators for two internal states $1$
and $2$, $\omega$ is the laser frequency, and
$A_k$ are Fourier components of the laser field. We ignore the interaction $\lambda_{22}$
of atoms in
the excited state, since we are interested in the absorption at small laser power,
when the density of excited atoms is negligible with respect to total density
$N$.

As discussed above, the interesting regime is that of a (nearly) Doppler
free two photon transition,
$\vec k=\vec k_1+\vec k_2$, $\omega=\omega_1+\omega_2$, where $\omega\simeq\omega_0$ and
$|\vec k|\simeq \lambda_{12}N/v_T\ll\omega_0/c$.

The absorption spectrum is expressed
through the two-particle correlation function by Kubo formula:
  \begin{eqnarray}\label{Kubo}
{\cal I}(\omega)&=&
{\rm Im} \int\limits^{\infty}_0
e^{i\omega t} \Big\langle\left[ \widehat C(t),\widehat C^+(0)\right]\Big\rangle dt
  \cr
\widehat C&=&\sum\limits_{p,k} A^*_k\widehat\eta_{p,k}\ ,\ \
  \widehat\eta_{p,k}=b_{p+k}a^+_p
\end{eqnarray}

  To evaluate the commutator in (\ref{Kubo}), we consider time
evolution of the operators: $i\partial_t
\widehat\eta_{p,k}=\left[\widehat\eta_{p,k}, {\cal H}_1+ {\cal H}_2
\right]$. In the situation of interest, when $\tau_{\rm
el}\gg\tau_{\rm exch}$, the commutator can be evaluated by the
RPA procedure\cite{RPA}. The resulting equations
for $\widehat\eta_{p,k}(t)$ are linear:
  \begin{equation}\label{RPAevolution}
i\partial_t \widehat\eta_{p,k}
=\Omega_{p,k}
\widehat\eta_{p,k}+\lambda_{12}n(p) \sum_{p'} \widehat\eta_{p',k}
  \end{equation}
where
  \begin{equation}
\Omega_{p,k}= \omega_0+\frac{(p+k)^2}{2m}-\frac{p^2}{2m}
+(\lambda_{12}-2\lambda_{11})N
  \end{equation}

To remind the reader how the RPA method works, consider the commutator
  \begin{eqnarray}\label{commutator}
& &\left[\widehat\eta_{p,k},
\sum\limits_{p_1+p_3=p_2+p_4}
a^+_{p_1} a_{p_2} b^+_{p_3} b_{p_4}\right]
=  \\
& &\sum\limits_{p_1+p+k=p_2+p_4}
a^+_p a^+_{p_1} a_{p_2} b_{p_4}
-
\sum\limits_{p_1+p_3=p+p_4}
a^+_{p_1} b^+_{p_3} b_{p_4} b_{p+k}
  \nonumber
  \end{eqnarray}
In the first term, according to the RPA procedure, we keep only
two kinds of contributions: with $p_2=p_1$, and with $p_2=p$.
Correspondingly, this gives
$N\widehat\eta_{p,k}+n(p)\sum_{p'}\widehat\eta_{p',k}$, which is the
part of (\ref{RPAevolution}) proportional to $\lambda_{12}$. The
second term in the commutator (\ref{commutator}) is of order of
the inverse volume of the system, and so it can be neglected.
(The first and second terms in (\ref{commutator}) are of
different magnitudes because there is only one particle $2$ in
the systems, while particles $1$ have finite density.)

Solving the system (\ref{RPAevolution}), and plugging the result in (\ref{Kubo}), one gets
the spectral power
  \begin{equation}\label{S(w)}
{\cal I}(\omega)= {\rm Im}
\sum\limits_k |A_k|^2 \frac{\Pi_0(k,\omega)}{1-\lambda_{12} \Pi_0(k,\omega)} ,
  \end{equation}
where
  \begin{equation}\label{P0}
\Pi_0(k,\omega)=\int {n(p)\over \omega-\Omega_{p,k}+i0}\ \frac{d^3p}{(2\pi\hbar)^3}
  \end{equation}
The integrals in (\ref{P0}) over components of $\vec p$ normal to $\vec k$
are evaluated explicitly, and the integral over longitudinal component of
$p$ is evaluated as a Cauchy integral, yielding an infinite (but quickly converging)
sum over poles in the complex $p$ plane\cite{OL}. The resulting spectrum
is shown in Fig.\ref{fig1} for several increasing densities.

To get the dispersion relation (\ref{dispersion})
one expands $\Pi_0$ in small $k$: $kv_T/\lambda_{12}N\ll1$.
The resulting dispersion (\ref{dispersion}), which is much stronger than that
for a free atom, is a collective mode effect\cite{Bashkin,Ruckenstein}.

To generalize the above calculation to the BEC situation, we introduce
$\widehat\eta_{0,k}=b_k\psi^*_0$, where $\psi_0$ is the condensate
amplitude. In the RPA calculation the operator $\widehat\eta_{0,k}$
is treated separately from other $\widehat\eta$'s. Now, the RPA
dynamics is the following:
  \begin{eqnarray}\label{RPA-Bose}
i\partial_t \widehat\eta_{p,k}
&=&\Omega_{p,k}\widehat\eta_{p,k}+
\lambda_{12}n(p)\left(\widehat\eta_{0,k}+
\sum_{p'} \widehat\eta_{p',k}\right) \cr
i\partial_t \widehat\eta_{0,k}
&=&(\Omega_{0,k}+\lambda_{11}N_c)
\widehat\eta_{0,k}+\lambda_{12}N_c \sum_{p'} \widehat\eta_{p',k}\ ,
  \end{eqnarray}
where $N_c=|\psi_0|^2$. Solving these equations yields the result of the form
(\ref{S(w)}),
  \begin{equation}\label{SBose(w)}
{\cal I}(\omega)= {\rm Im}
\sum\limits_k |A_k|^2 \frac{\Pi(k,\omega)}{1-\lambda_{12} \Pi(k,\omega)} ,
  \end{equation}
where
  \begin{equation}\label{P1}
\Pi(k,\omega)=\frac{N_c}{\omega-\Omega_{0,k}+(\lambda_{12}-\lambda_{11})N_c+i0} +
\Pi_0(k,\omega)
  \end{equation}
The spectrum at $k=0$ consists of two sharp lines, with frequencies given by
(\ref{2frequencies}).
The physics of the doubling of the excitation below $T_{\rm
BEC}$ is similar to that of the second sound. The excited
particle can be in two different dynamical states, either having
energy of order $T$ and participating in momentum exchange
processes, or being at rest. These two states are coupled by the
transitions provided by momentum exchange processes (see
(\ref{RPA-Bose})), in which the excited particle is transferred
between the thermal and stationary states by exchanging with a
condensate particle. The rate of such (coherent) transitions
determines the frequency splitting in (\ref{2frequencies}).

Relative strength of the spectral components given by
(\ref{SBose(w)}) depends on the condensate fraction $N_c/N$. At
$N_c\ll N$ the peak due to the states with thermal  energies  is
much stronger than the other peak, which is due to atoms at
rest. Because of that, we call the two peaks ``normal'' and
``condensate'', respectively. At $N_c\to N$ the strengths of the
two peaks become equal, and at $T\ll T_{\rm BEC}$ the peaks
approach each other. This behavior is displayed in
Figs.\ref{fig2},\ref{fig3}.
\noindent{\it Conclusion:} We described phenomena in optical
excitations resulting from cold collisions of the excited atom
with other atoms of the gas. The main feature that makes the
effect of collisions an interesting problem is that, during the
optical transition, the excited atom spends a long time in the
superposition of the ground and excited state. Because of many
collisions that occur during this time, and because of their
special coherent momentum--exchange character, the optical
excitation represents a collective mode. The collective nature
of the excitation is manifest in the frequency shift and in the
linewidth narrowing. Below the BEC transition, due to the
exchange between the condensate and the normal component, the
number of modes doubles, similar to the first and second acoustic
modes. By using the RPA method, we
calculate the temperature--dependent frequencies of the doublet
and the oscillator strengths.
\acknowledgements
\noindent
We are grateful to Tom~Greytak for proposing the problem, and to
Dan~Kleppner, Tom~Killian, and Wolfgang~Ketterle for many useful
discussions. We acknowledge NSF support under award DMR94-00334.

\end{document}